\documentclass[runningheads]{llncs}
\usepackage[T1]{fontenc}
\usepackage{graphicx}
\usepackage{epstopdf}
\usepackage{booktabs}
\usepackage{array}
\usepackage{amsmath}
\usepackage{amssymb} 
\usepackage{subcaption}
\usepackage[labelfont=bf]{caption}
\usepackage{hyperref} 
\usepackage{multirow}
\usepackage{multicol}
\usepackage{marvosym}

\newcolumntype{M}[1]{>{\centering\arraybackslash}m{#1}}
\begin{document}
\title{Estimation of Time-to-Total Knee Replacement Surgery}
\author{Ozkan Cigdem\inst{1}\textsuperscript{(\Letter~\texttt)}  \and Shengjia Chen\inst{1} \and Chaojie Zhang\inst{1}  \and Kyunghyun Cho\inst{2} \and Richard Kijowski\inst{1} \and Cem M. Deniz\inst{1}}
\authorrunning{Ozkan Cigdem et al.}
%
\institute{\inst{} Department of Radiology, New York University Grossman School of Medicine, New York, USA \\ \email{ozkan.cigdem@nyulangone.org} \and \inst{} Center of Data Science, New York University, New York, USA
}

%
%
\maketitle              
\begin{abstract}
A survival analysis model for predicting time-to-total knee replacement (TKR) was developed using features from medical images and clinical measurements. Supervised and self-supervised deep learning approaches were utilized to extract features from radiographs and magnetic resonance images. Extracted features were combined with clinical and image assessments for survival analysis using random survival forests. The proposed model demonstrated high discrimination power by combining deep learning features and clinical and image assessments using a fusion of multiple modalities.  The model achieved an accuracy of 75.6\% and a C-Index of 84.8\% for predicting the time-to-TKR surgery. Accurate time-to-TKR predictions have the potential to help assist physicians to personalize treatment strategies and improve patient outcomes.

\keywords{Knee osteoarthritis \and Survival Analysis \and Artificial Intelligence \and Deep Learning \and Random Survival Forest}
\end{abstract}
\section{Introduction}
Osteoarthritis is a prevalent joint disease, posing a significant global health challenge and resulting in physical disability \cite{Kellgren1957}. Amongst the various types, knee osteoarthritis (KOA) emerges as the most common form of arthritis, profoundly affecting the quality of life for millions worldwide by inducing pain, mobility constraints, and disability\cite{Altman1986}. KOA directly affects about 10\% of men and 13\% of women aged 60 and above. Despite lacking a definitive cure for KOA, total knee replacement (TKR) surgery emerges as a plausible intervention during advanced disease stages \cite{Felson1998}. Predicting the time until TKR is essential for identifying patients at higher risk of rapid KOA progression and offering prognostic guidance to those considering TKR before potential future health decline \cite{Jamshidi2021}. 

Predicting time-to-TKR involves using both symptoms reported by patients and findings from imaging modalities \cite{Jamshidi2021,Heisinger2020,Mahmoud2023}. Among these modalities, radiography and magnetic resonance imaging (MRI) stand out as the most commonly employed techniques for evaluating KOA \cite{Wang2022,KNOAP2020_Challenge,Sneag2023,Cigdem2023Review}. Features on radiographs and MRI scans, coupled with quantitative and semi-quantitative assessments of imaging, have demonstrated associations with KOA progression, including an elevated risk for TKR \cite{Leung2020,Mahmoud2023,Tolpadi2020,Panfilov2022_Progression}. However, accurately determining the time-to-TKR is a complex task affected by multiple factors. It depends not only on the progression of structural disease but also on patient-specific factors such as personal preferences, financial limitations, the presence of other medical conditions, and overall health conditions. These factors introduce a level of variability that complicates the prediction of patient timelines \cite{Wetstein2022,Adeoye2022}. To overcome the complexities of various factors influencing time-to-TKR, advanced tools are necessary for accurate prediction.

Deep learning (DL) shows promise in overcoming the challenges of predicting the surgery time or brain age \cite{Hedayati2023,Yang_Yanwu2022}. With enough training data, a DL model can automatically identify crucial features from radiographs and MRIs that are associated with the progression of KOA \cite{Rajamohan2023,Tolpadi2020,Panfilov2022_Progression}. Survival prediction models estimate patient survival likelihood, crucially incorporating also the right-censored data representing event-free duration \cite{Suresh2022}. Neglecting censored data may yield biased results, underscoring its importance in accurate predictions. Our study introduced a multi-modal approach that utilized right-censored data, clinical variables, quantitative and qualitative measurements from radiographs and MRIs, as well as DL features extracted from these modalities to predict the time-to-TKR over a 9-year follow-up period. We hypothesize that integrating clinical variables, quantitative and semi-quantitative assessments from radiographs and MRIs, and DL features into survival models will result in more accurate estimations of time-to-TKR compared to models using only DL features.

\section{Methods}
\subsection{Study cohort}
The study utilized knee data from the Osteoarthritis Initiative (OAI) \cite{Lester2008_OAI} publicly accessible database. The OAI database contains clinical variables, radiographs, MRI exams, and radiograph and MRI quantitative and semi-quantitative image assessment measurements for 4,796 subjects aged 45 to 79 with or at risk for KOA, evaluated at baseline and follow-ups at 12, 18, 24, 30, 36, 48, 60, 72, and 96 months. The OAI received ethical approval from the Internal Review Boards at the University of California at San Francisco and each individual clinical recruitment site. All participants provided written informed consent.

The study cohort in the OAI was evaluated with longitudinal radiographs and MRI exams consisting of sagittal fat-suppressed intermediate-weighted turbo spin-echo (TSE) and sagittal fat-suppressed three-dimensional dual-echo in steady state (DESS) sequences. Out of 4,796 subjects from the OAI database, 547 subjects underwent TKR during the 9-year follow-up period. Each subject may have undergone TKR in either one or both knees (163 with only left knee, 168 with only the right knee, and 108 with both knees). Both knees of the same patients were used as separate data points to enhance model robustness. The clinical variables, radiographs, MRI exams, and radiographic and MRI quantitative and quantitative image assessment measurements at follow-up time points were used for each knee undergoing TKR with the time of follow-up considered year 0 for estimating time-to-TKR.

\begin{figure}[htb!]
\centering
\includegraphics[width=1.0\linewidth, height=0.3\textheight,keepaspectratio]{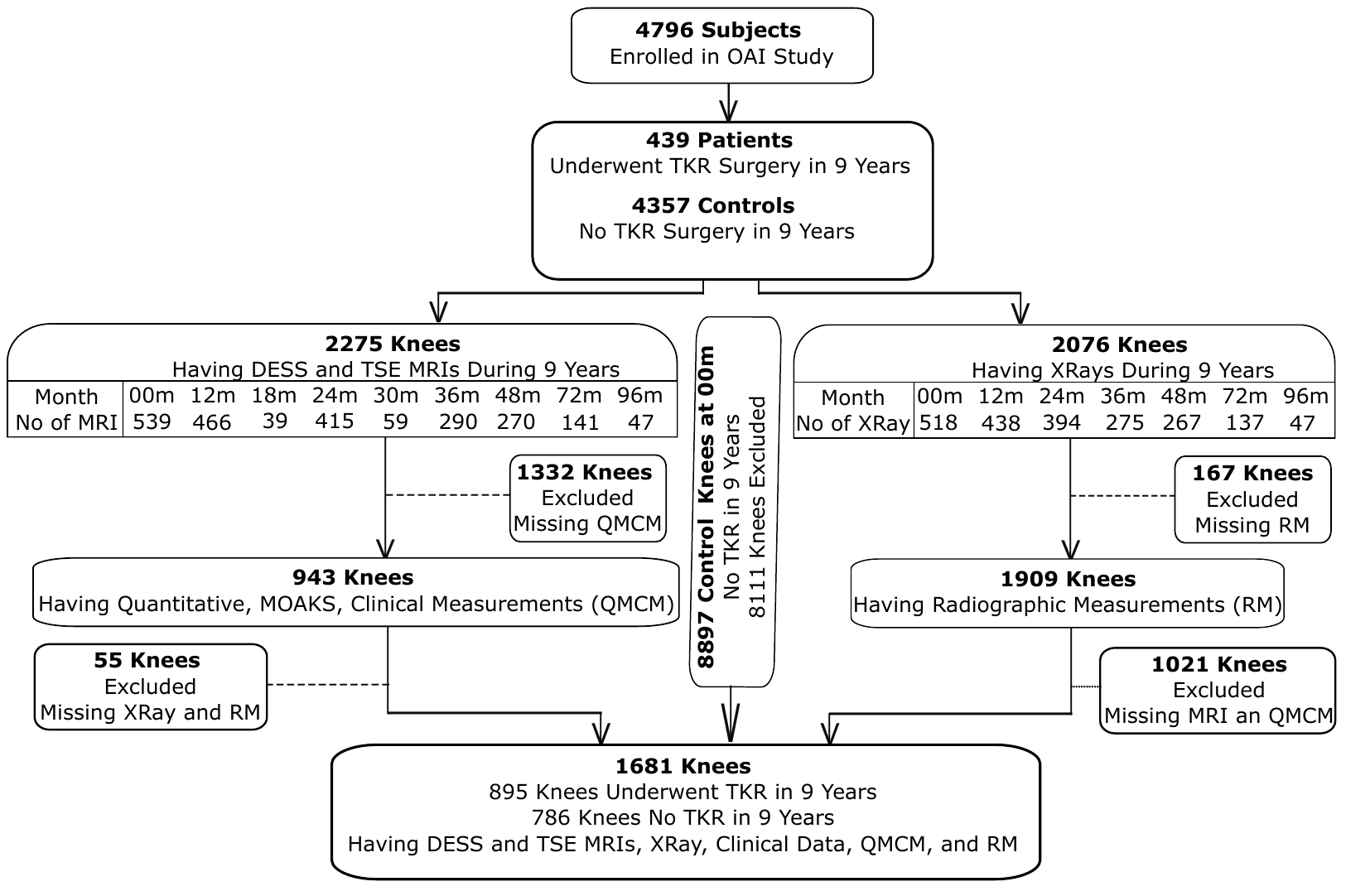}
\caption{After reviewing clinical, radiograph, and MRI data along with clinical, quantitative, and semi-quantitative image assessment measurements, 1681 knee data from the OAI database were identified. Knees:knee images.}
\label{DatasetSelection}
\end{figure}

\begin{table}[!h]
\centering
\caption{Baseline gender and age of subjects in the OAI study cohort. F: female, M: male, Mean: mean of age, std: standard deviation, Train/Val./Test: number of training, validation, and test groups.}
\label{demographicInfo}
\resizebox{\textwidth}{!}{
\begin{tabular}{M{3cm}|M{2cm}|M{2cm}|M{2cm}|M{2cm}}
\toprule
Database & No. of & Gender: & Age & Age \\
Train/Val./Test & Images & Patients & Mean $\pm$ std & Range \\ \midrule
 & TKR & M: 109 & 64.2 $\pm$ 9.3 & 46-77 \\
OAI & 895 & F: 170 & 62.4 $\pm$ 8.1 & 45-76 \\ \cline{2-5}
1239/172/270 & Control (right-censored) & M: 322 & 62.1 $\pm$ 9.3 & 45-79 \\
 & 786 & F: 422 & 61.9 $\pm$ 9.3 & 45-79 \\ \bottomrule
\end{tabular}
}
\end{table}

A total of 1,681 knees (895 knees with TKR within 9 years and 786 knees as right-censored controls)  have complete clinical variables, radiograph and MRI quantitative and semi-quantitative image assessment measurements, and MRI semi-quantitative image assessment measurements using the MRI osteoarthritis knee score (MOAKS) system \cite{Hunter2011}. The study cohort identification is summarized in Fig. \ref{DatasetSelection}. The dataset was partitioned into 1,239 training, 172 validation, and 270 test data. The training, validation, and test data splits were done at the subject level so that all follow-up data associated with the same subject was included in a single split. To ensure consistency, the same data splits were used for training, validation, and testing of both unsupervised and supervised models. The details of the study cohort and data splits are presented in Table \ref{demographicInfo}.

\begin{figure}[h!]
\centering
\includegraphics[width=\linewidth,height=0.4\textheight,keepaspectratio]{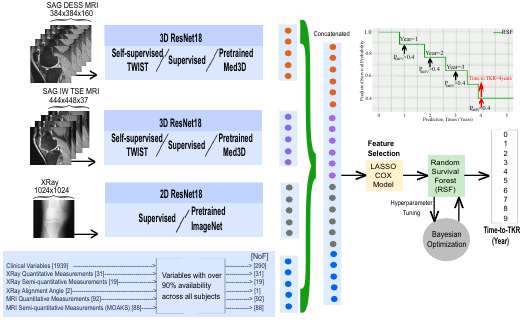}
\caption{The flowchart of the proposed approach.}
\label{Flowchart}
\end{figure}

\subsection{Proposed Model}
\paragraph{\textbf{Self-supervised Pre-training.}}
The ImageNet pretrained Resnet18 model was used to extract features from radiographs. Two separate 3D ResNet18 were trained on the TSE and DESS MRI sequences using a self-supervised framework, Twin Class Distribution Estimation (TWIST) \cite{wang2023self}. The TWIST framework was used to extract representative features from unlabeled 3D knee MR images. The framework employed a siamese network to generate twin class distributions of two augmented MR images. A combination of three loss terms was introduced to encourage the model to extract distinct features of different MR images. (1) The consistency term enforced the class distributions of two augmented views to be consistent. (2) The sharpness term enforced each sample distribution to be sharp, making each sample have a deterministic assignment. (3) The diversity term enforced different samples to be diversely distributed to different classes. Trained TWIST models were used to extract features from the study cohort involving TKR patients and right-censored controls. 

\paragraph{\textbf{Supervised Training.}} Two separate 3D ResNet18 models were trained using the TSE and DESS sequences and a 2D ResNet18 model was trained using the radiographs. For image modality, the labels representing time-to-TKR were mapped to a normal distribution (discretized into 30 bins) with a variance of 4 and a mean equal to the label \cite{Yang_Yanwu2022,Chen_S2019}. The model was trained using the Kullback–Leibler divergence loss, which ensured that the output of the model had a distribution similar to the normal distribution of the labels. Predictions were calculated based on the area under the predicted distribution. The best model for each MRI sequence was selected based on the highest accuracy achieved on the validation data \cite{Hedayati2023}. Trained DL models were used to extract the features from the TKR patients.

\paragraph{\textbf{Feature Selection and Ensemble Model Training.}} DL features were extracted from the output of the last pooling layer. Of the 1,939 baseline clinical variables, 290 were available for over 90\% of the subjects in the OAI database. All available quantitative and semi-quantitative image assessment measurements were utilized. Missing measurements in the dataset were imputed using the mean of non-missing values for quantitative data and the mode of non-missing values for categorical variables. A least absolute shrinkage and selection operator (Lasso) method was applied to the Cox regression model for feature selection \cite{Jamshidi2021}. A random survival forest (RSF) model \cite{Suresh2022} was used for time-to-TKR prediction. The same training and validation data cohorts utilized in the DL models were used in the RSF model \cite{Suresh2022}. The RSF model's output consisted of predicted survival probabilities, indicating the likelihood of not undergoing TKR surgery, for each subject over a 9-year period. When the survival probability fell below the threshold of 0.4, the time-to-TKR prediction timing corresponded to the latest year in which it surpassed the threshold. The proposed model's flowchart is presented in Fig. \ref{Flowchart}. To ensure a fair comparison between the proposed model and the existing literature, the algorithms used in previous studies were applied to the datasets used in this research. The Wilcoxon signed-rank test was used to evaluate differences in performance between the different models. The source code for this study is available at \href{link will be provided }{(link will be provided)}.

\section{Experiments and Results}
Table \ref{PredictionResults_MRI} compares the performances of various models using both DL model features extracted from radiographs and the TSE and DESS sequences and incorporating them to the clinical variables, radiograph and MRI quantitative and semi-quantitative image assessment measurements for time-to-TKR estimation. The RSF model utilizing self-supervised DL model features extracted from the DESS and TSE MRI sequences exhibited comparable accuracy (42.2\%) but achieved a superior C-Index (69.8\%) compared to the RSF model utilizing pre-trained Med3D DL model features extracted from the DESS and TSE sequences, which attained an accuracy of 41.5\% and a C-Index of 56\% \cite{Chen2019Med3DTL}. The RSF model with self-supervised DL model features extracted from radiographs and the DESS and TSE MRI sequences improved the accuracy of time-to-TKR estimation (49.3\%) compared to self-supervised  DL model features extracted from DESS and TSE MRI sequences alone (42.2\%). Likewise, the RSF model with Lasso Cox feature selection method combining clinical variables, radiograph and MRI quantitative and semi-quantitative image assessment measurements, and the concatenated self-supervised DL model features extracted from the DESS and TSE sequences and ImageNet pretrained DL model features extracted from radiographs had the highest estimation accuracy (75.6\%) and C-Index (84.8\%) for predicting time-to-TKR, which was significantly higher than the estimation accuracy (49.3\%) and C-Index (79.8\%) of the combination of these DL model features extracted from radiographs and the DESS and TSE sequences alone.

\begin{figure}[t!]
\centering
\begin{subfigure}{0.45\textwidth}
    \centering
    \includegraphics[width=\linewidth]{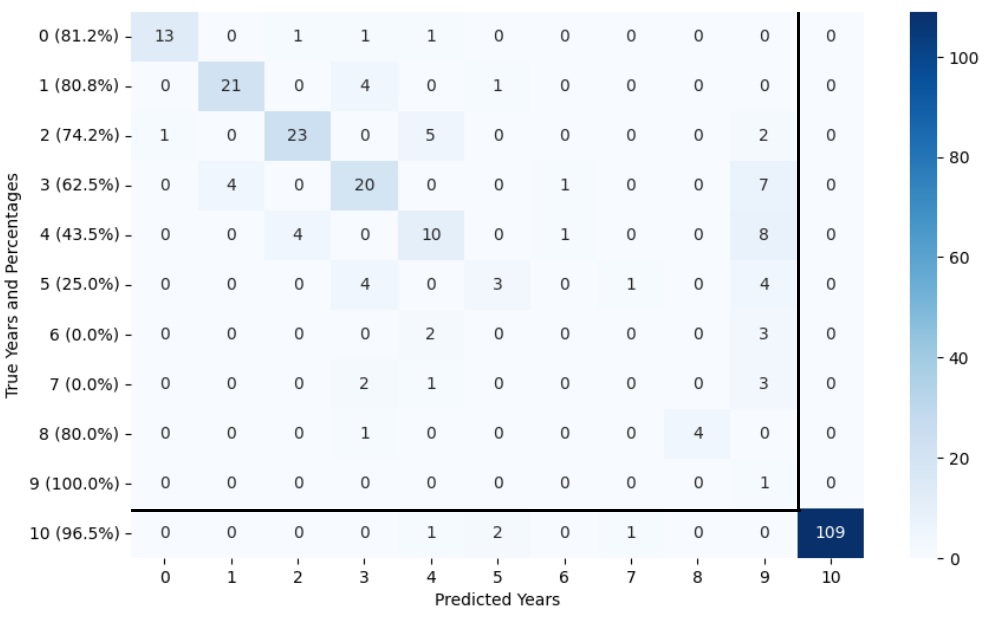}
    \caption{}
    \label{CM_MRISubset}
\end{subfigure}
\begin{subfigure}{0.45\textwidth}
    \centering
    \includegraphics[width=\linewidth]{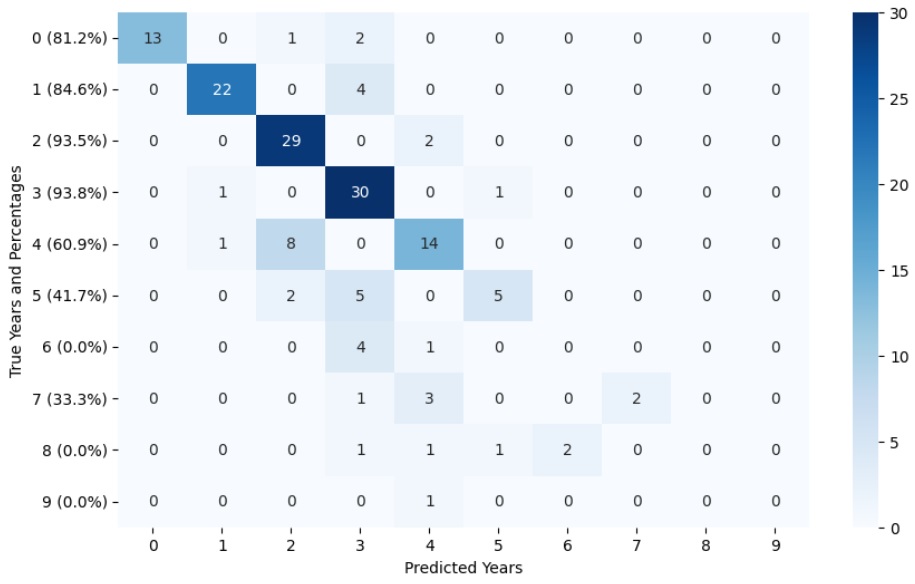}
    \caption{}
    \label{CM_MRIXraySubset}
\end{subfigure}
\caption{ Confusion matrices for $\pm 1$ year estimation ($|y-\hat{y}|\leq 1$) for the following models from Table \ref{PredictionResults_MRI}:\\
a) \textsuperscript{\textdagger}(Self-supervised(DESS+TSE)+Pretrained-ResNet(XRay)+Clinical+Measurements) with data including both the subjects underwent TKR in 9-years and right-censored controls),\\
b) \textsuperscript{\textdagger}(Supervised(DESS+TSE+XRay)+Clinical+Measurements) with only the subjects underwent TKR in 9-years.\textbf{\textsuperscript{\textdagger}}: LASSO Cox feature selection method was used before RSF model.
}
\label{ConfusionMatrices}
\end{figure}

The RSF model with Lasso Cox feature selection method combining clinical variables, radiograph and MRI quantitative and semi-quantitative image assessment measurements, and the concatenated supervised DL model features extracted from radiographs and the DESS and TSE sequences had better estimation accuracy (73.2\%) and C-Index (76.2\%) for predicting time-to-TKR, which was significantly higher than the estimation accuracy of (63.7\%) and C-Index (68.9\%) of the concatenated supervised DL model features extracted from radiographs and the DESS and TSE sequences alone. The confusion matrices for $\pm 1$ year estimation ($|y-\hat{y}|\leq 1$) are shown in Fig. \ref{ConfusionMatrices}.

\begin{table}[!h]
\centering
\caption{SSL: Self-supervised learning, ACC(\%): accuracy (number of correctly predicted subjects/all subjects), AUC(\%): mean AUC over 9 years, IBS: mean integrated brier score over 9 years, C-In.(\%): concordance index, Measurements: MRI osteoarthritis knee score and quantitative MRI image assessments, and quantitative, semi-quantitative, and alignment measurements of radiographs, NoF: number of features, \textbf{\textsuperscript{\textdagger}}: LASSO Cox feature selection method was used before RSF model, $+$: concatenation.}
\label{PredictionResults_MRI}
\resizebox{\textwidth}{!}{
\begin{tabular}{c|c|ccccc|cccccc}
\toprule
 & & \multicolumn{5}{c}{RSF Model Input}  & \multicolumn{6}{|c}{Evaluation Metrics}  \\ \midrule
Model &\textbf{\textsuperscript{\textdagger}} & DESS & TSE & XRay & Clinical & Measurements & NoF & ACC & C-In. & AUC & IBS & p \\ 
\midrule
Med3D & &$\checkmark$ & $\checkmark$ &  &  &  & 1024 & 41.5 & 56.0 & 71.9 & 0.118 & 0.002 \\
\midrule
 \multirow{4}{*}{SSL TWIST} & &$\checkmark$ & $\checkmark$ &  &  &  & 512 & 42.2 & 69.8 & 76.1 & 0.106 & 0.002 \\
 & &$\checkmark$ & $\checkmark$ & $\checkmark$ &  &  & 1024 & 49.3 & 73.8 & 82.1 & 0.095 & $*$ \\
 & &$\checkmark$ & $\checkmark$ & $\checkmark$ & $\checkmark$ & $\checkmark$ & 1545 & 74.4 & \textbf{85.0} & 93.8 & 0.058 & $<0.001$ \\
 & \checkmark &$\checkmark$ & $\checkmark$ & $\checkmark$ & $\checkmark$ & $\checkmark$ & \textbf{113}& \textbf{75.6} & 84.8 & \textbf{94.9} & \textbf{0.055} & $<0.001$ \\ \midrule
- & \checkmark & &  &  & $\checkmark$ & $\checkmark$ & 90 & 73 & 82.3 & 94 & 0.06 & $<0.001$ \\
\midrule
\multirow{3}{*}{Supervised ResNet18} & &$\checkmark$ & $\checkmark$ & $\checkmark$ &  &  & 1024 & 63.7 & 68.9 & 78.7 & 0.105 & $*$ \\
 & &$\checkmark$ & $\checkmark$ & $\checkmark$ & $\checkmark$ & $\checkmark$ & 1545 & 70.7 & 71.8 & 78.4 & 0.098 & $<0.001$ \\
 & \checkmark &$\checkmark$ & $\checkmark$ & $\checkmark$ & $\checkmark$ & $\checkmark$ & 91 & 73.2 & 76.2 & 85.4 & 0.09 & $<0.001$ \\
\bottomrule
\end{tabular}
}
\end{table}

\begin{table}[hbt!] 
\centering
\caption{Comparing the performance of existing methods using the subject cohort from this paper. The differences in the number of features arise from two factors: (1) missing values for more than 50\% of knees, and (2) enrollment variables (race, sex, history of knee arthroscopy) being identical across all visits. ACC: accuracy (number of correctly predicted subjects/all subjects), C-In.(\%): concordance index, NoF*: number of features used in the original study, NoF: number of features, Radiographic: quantitative, semi-quantitative, and alignment measurements of radiograph images.}
\resizebox{\textwidth}{!}{
\begin{tabular}{l l c c c c }
\hline
Research Work & RSF Model Input (NoF*) & \multicolumn{1}{l|}{NoF} & \multicolumn{1}{l|}{ACC(\%)} & \multicolumn{1}{l|}{C-Index(\%)} & \multicolumn{1}{l|}{P-value}\\ \hline
Jamshidi et al. \cite{Jamshidi2021}         & MOAKS+Radiographic+Clinical (3) & 3  & 56.7 & 71.5 & $<$0.001  \\ 
Heisinger et al. \cite{Heisinger2020}     & Radiographic+Clinical (14)                                      & 14  & 62.1 & 63.4 & $<$0.001  \\
Mahmoud et al. \cite{Mahmoud2023}          & Radiographic+Clinical (45)                                      & 31  & 45.2 & 62.7 & $<$0.001  \\  
Liu et al. \cite{Qiang_Liu2022}            & Radiographic+Clinical (9)                                       & 6   & 60 & 73.5 & $<$0.001 \\ 
Hu et al \cite{Hu2023_deepKOA}             & DLmodel\textsubscript{TSE}                                      & 256 & 40 & 67.5 & $<$0.001  \\ 
\multirow{1}{*}{\textbf{Our Study}}     & \textbf{Proposed Model\textsubscript{Xray+TSE+DESS}}  & \textbf{113}    & \textbf{75.6} &\textbf{84.8} & $*$ \\ \hline 
\end{tabular}}
\label{ReferenceComparison}
\end{table}

\subsection{Ablation Study}
The performance of utilizing clinical variables together with quantitative and semi-quantitative assessments from radiographs and MRI scans, DL features extracted from these imaging modalities, and their combination were analyzed using three different models: the Cox proportional hazards model, discrete-time Generalized Linear Model (GLM), and RSF model. The experiments were conducted with the study cohort, and among them, the RSF model yielded the most favorable results.

\begin{table}[hbt!] 
\centering
\caption{The performance of using different models for prediction of time-to-TKR,  \textbf{\textsuperscript{\textdagger}}: LASSO Cox feature selection method was used before RSF model}
\begin{tabular}{c|ccccc|ccc}
\toprule
\textbf{\textsuperscript{\textdagger}} & DESS & TSE & XRay & Clinical & Measurements & Cox & GLM & RSF \\ 
\midrule
& &  &  & $\checkmark$ & $\checkmark$ & 0.901 & 0.898 & 0.94 \\
& $\checkmark$ & $\checkmark$ & $\checkmark$ &  &  & 0.809 & 0.811 & 0.821 \\
& $\checkmark$ & $\checkmark$ & $\checkmark$ & $\checkmark$ & $\checkmark$ & 0.904 & 0.895 & 0.938 \\
$\checkmark$ & $\checkmark$ & $\checkmark$ & $\checkmark$ & $\checkmark$ & $\checkmark$ & 0.935 & 0.936 & 0.949 \\
\bottomrule
\end{tabular}
\label{AblationStudy}
\end{table}

\section{Discussion}
Our study demonstrated that integrating various clinical variables, along with quantitative and semi-quantitative assessments from radiographs and MRI scans, as well as DL features extracted from these imaging modalities, into predictive models resulted in superior accuracy in estimating the time-to-TKR compared to models analyzing DL features individually. The most effective model utilized the whole study cohort (TKR patients and right-censored data) with 113 relevant clinical variables, quantitative and semi-quantitative assessments from radiographs and MRI scans, and self-supervised DL model features trained on data from these modalities. The performance of using self-supervised pretrained DL model provided better prediction performance on estimation of time-to-TKR compared to that of supervised DL model using only the subjects underwent TKR in 9 years time frame. The use of right-censored data may help capture the entire distribution of time-to-TKR within the study cohort and has the potential to enhance the model's generalizability by incorporating all available information for predicting the true risk of requiring TKR in the future. 

Application of machine learning models in predicting TKR by leveraging clinical variables and quantitative and semi-quantitative assessments from radiographs and MRI scans have been investigated in the existing literature \cite{Jamshidi2021,Heisinger2020,Hu2023_deepKOA,Mahmoud2023,Qiang_Liu2022}. Compared to previous approaches, our study incorporated DL features extracted from baseline radiographs and MRI into ML models to predict time-to-TKR. Our model had higher estimation accuracy and C-index than previous studies \cite{Jamshidi2021,Heisinger2020,Mahmoud2023,Qiang_Liu2022}, which is likely due to the inclusion of a much larger number of clinical variables and radiograph and MRI image assessment measurements.

In this study, the data from the OAI database which is primarily composed of older, overweight, and Caucasian subjects was used. Thus, model generalizability to more age, body mass index, race, and ethnic diverse subject populations needs to be further investigated. Additionally, the decision to undergo a TKR can be affected by multiple factors other than the severity of KOA such as personal preferences, pain tolerance, and access to healthcare, and these variables were not available in the OAI databases for use in our study.

\section{Conclusion}
Accurately predicting the time-to-TKR is a challenging task due to various factors associated with the surgery decision. The use of an ensemble model for survival analysis, incorporating clinical variables, image assessment measurements, and MRI sequence as well radiograph features extracted from unsupervised trained DL model, represented promise in estimating the time-to-TKR. 

%
%
%

\bibliographystyle{splncs04}
\bibliography{MICCAI}
\end{document}